\documentclass[aps,pra,reprint,groupedaddress,longbibliography]{revtex4-2}
\usepackage{graphicx,siunitx}
\usepackage{hyperref}
\hypersetup{colorlinks=true}
\usepackage{braket}
\usepackage{float}
\usepackage{amsmath}
\usepackage{xcolor,soul}

\newcommand\Ca{$^{40}\text{Ca}^+$}
\newcommand{\nth}{\braket{n_{\mathrm{th}}}}
\newcommand{\ncoh}{\braket{n_{\mathrm{coh}}}}
\newcommand{\designSpeed}{100~m/s}
\newcommand{\averageSpeed}{82(2)~m/s}
\newcommand{\speed}{251(6)~m/s}

\usepackage[normalem]{ulem}

\begin{document}

\title{Characterization of Fast Ion Transport via Position-Dependent Optical Deshelving}

\author{Craig R. Clark} 
\email[]{craig.clark@gtri.gatech.edu}
\author{Creston D. Herold, J. True Merrill, Holly N. Tinkey, Wade Rellergert, Robert Clark, Roger Brown, Wesley D. Robertson, Curtis Volin, Kara Maller, Chris Shappert, Brian J. McMahon}
\author{Brian C. Sawyer}
\email[]{brian.sawyer@gtri.gatech.edu}
\author{Kenton R. Brown}
\email[]{kenton.brown@gtri.gatech.edu}
\affiliation{Georgia Tech Research Institute, Atlanta, Georgia 30332, USA}

\begin{abstract}
Ion transport is an essential operation in some models of quantum information processing, where fast ion shuttling with minimal motional excitation is necessary for efficient, high-fidelity quantum logic. While fast and cold ion shuttling has been demonstrated, the dynamics and specific trajectory of an ion during diabatic transport have not been studied in detail. Here we describe a position-dependent optical deshelving technique useful for sampling an ion's position throughout its trajectory, and we demonstrate the technique on fast linear transport of a \Ca{} ion in a surface-electrode ion trap. At high speed, the trap's electrode filters strongly distort the transport potential waveform. With this technique, we observe deviations from the intended constant-velocity (100~m/s) transport: we measure an average speed of {\averageSpeed{}} and a peak speed of {\speed{}} over a distance of 120~$\mu$m.
\end{abstract}

\maketitle

Systems of trapped atomic ions represent some of the most promising platforms for quantum information processing, benefiting from long qubit coherence times and from the highest operational fidelities demonstrated to date \cite{clark_2021,srinivas_2021,ryan-anderson_implementing_2022}. In the QCCD ion-trap architecture \cite{kielpinski_architecture_2002}, ions are transported between various regions within the processor to reconfigure which ions can interact at any given step of an algorithm. Such an architecture demands finely tuned control of ion shuttling in order to implement reconfigurations as quickly as possible without also degrading the fidelities of subsequent logic gates \cite{clark_2021, ballance_2016, sorensen_2000}. Performing logic operations on the ions at the same time as they are transported within the trap also requires well characterized and reproducible trajectories for success \cite{de_clercq_parallel_2016, tinkey_transport-enabled_2022}.

Recent work has shown that ion transport contributes substantially to the latency of current QCCD systems \cite{pino_2021}.
Rapid linear ion transport has been demonstrated previously \cite{Bowler2012,walther_controlling_2012, sterk_closed-loop_2022} but has not been widely adopted likely due to the demands it places on waveform control hardware and on motional characterization and calibration. Transport at slower speeds has also been achieved through more complicated structures such as junctions of linear sections \cite{hensinger_t-junction_2006,blakestad_near-ground-state_2011,wright_reliable_2013,decaroli_design_2021,burton_transport_2022}.
While prior results have demonstrated transport between two locations with sub-quanta motional excitation, the ion trajectory during transport has not been measured. In fact, all prior reported transport speeds assumed an average speed calculated from the distance between static well positions and the designed playback speed of the trapping voltage waveforms. Some experiments have proved that an ion was transported to an intended location, e.g. by applying a focused laser pulse at that location \cite{Bowler2012,sterk_closed-loop_2022}; however, those experiments did not verify the exact time of arrival.

In this work, we present a method to measure the location of an ion throughout the entire arc of its transport, which enables us to reliably extract both instantaneous and average velocity during transport. This is achieved by measuring the probability for the ion to undergo a spontaneous irreversible transition from a metastable excited state to a ground state when illuminated by a laser beam having a spatial intensity gradient. Similar ideas have been explored in the context of neutral atoms in optical cavities \cite{puppe_single-atom_2004,du_precision_2013}, as well as in the field of ultrafast physics \cite{maiuri_ultrafast_2020}. The method presented here contrasts with previous schemes employing Fourier-limited coherent optical interactions to extract in-flight Doppler shifts~\cite{tinkey_transport-enabled_2022, de_clercq_parallel_2016}.

Here, we leverage the technique to show that an ion's trajectory deviates significantly from a simple constant-velocity path as the trap's electrode filters distort the applied potentials. Although the naive trajectory would have a constant speed of \designSpeed{}, corresponding to a displacement of 120~$\mu$m (two electrodes) in 1.2~$\mu$s, the ion actually achieves a maximum speed of \speed{} in the middle of its path.  These trajectory deviations lead to a large coherent displacement of the ion's motional state. We superimpose an additional compensating sinusoidal potential onto the waveform to remove this displacement \cite{Bowler2013,sterk_closed-loop_2022} and achieve a final transport-induced excitation of 0.7(2) quanta. Alternatively, it is possible in principle to pre-compensate the waveforms to account for the high-frequency attenuation of the trap electrode filters, but recent work highlights the technical difficulty of this approach in the context of fast ($\gtrsim 10$ m/s) transport~\cite{todaro_scalable}. 

\begin{figure}
\centering
\includegraphics[width=0.49\textwidth]{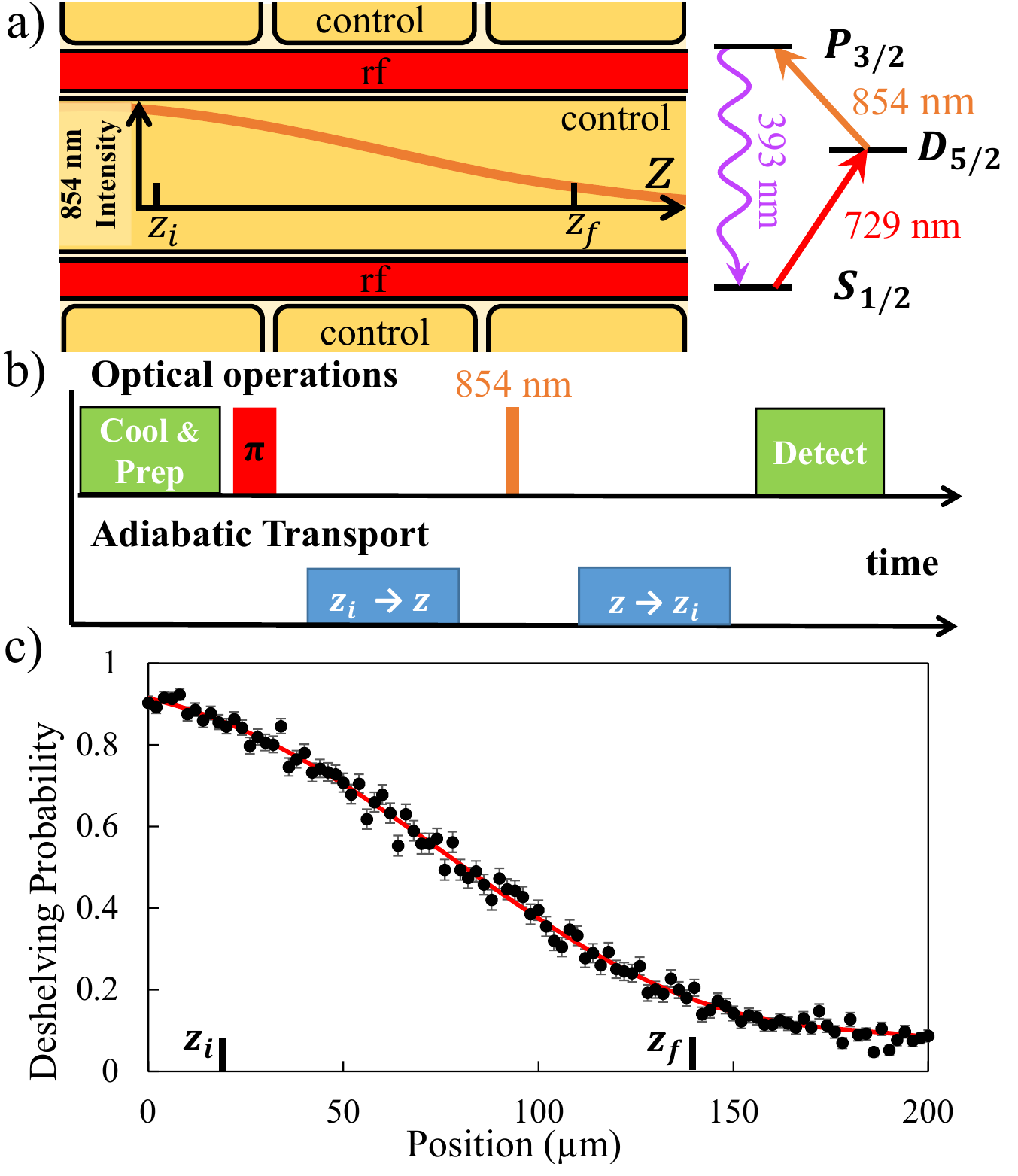}%
\caption{\label{fig:diagram} (color online) (a) (Left) Schematic of a 180~$\mu$m portion of the BGA trap (six out of 42 control electrodes) with a graph of the nominal 854~nm repumper intensity profile superimposed. Between the rows of control electrodes are two rf rails and a central grounded control electrode. The initial and final positions of the ion for the characterized transport are $z_i$ and $z_f$. (Right) Energy level diagram of \Ca{} ion showing the shelving transition driven by 729~nm light, the deshelving transition driven by 854~nm light, and spontaneous decay through emission of 393~nm light. (b) Ion trajectory calibration sequence. An ion in a stationary potential at $z_i$ is cooled and prepared in the $S_{1/2}$ level and then shelved into the $D_{5/2}$ level with a resonant 729 nm $\pi$-pulse (labeled $\pi$ in the timeline). The ion is adiabatically transported to another position along the trap axis, and the 854 nm beam is pulsed for 200 ns, thereby deshelving a portion of the $D_{5/2}$ population. This deshelving fraction is in one-to-one correspondence with the local beam intensity. The ion is shuttled adiabatically back to its initial location $z_i$. (c) Deshelving probability as a function of ion position (black points) with error bars representing the 68\% confidence interval assuming binomial statistics. The solid line represents a polynomial fit to the data.}
\end{figure}

Our experimental system employs a GTRI/Honeywell Ball Grid Array (BGA) trap \cite{Guise2015} with a 60~$\mu$m ion height, shown schematically in Fig.~\ref{fig:diagram}a, mounted in a room-temperature ultrahigh vacuum chamber and confining a single \Ca{} ion. The trap radiofrequency (rf) electrode is driven at 55~MHz and realizes approximate radial frequencies of 3.7 and 4.2~MHz. We determine the control electrode voltages necessary to produce axial confinement at different positions along the trap symmetry axis using an in-house boundary element method electrostatic solver \cite{charles_doret_controlling_2012}. We employ NIST-designed digital-to-analog converters (DACs) \cite{Bowler2013} to vary the potentials on the axial control electrodes. Each control electrode's potential is filtered with a two-pole low-pass filter having a bandwidth of 608~kHz ~\footnote{The trap filters consist of two low-pass stages: the first is a series inductor with shunt capacitor to ground, and the second is a series resistor with a shunt capacitor tuned to damp voltage `ringing' from the first stage}. The axial trap frequency (parallel to the axis of symmetry) is approximately 2.2~MHz. The axial heating rate for a stationary potential is 210~quanta/s.

For Doppler cooling we use a 397~nm laser beam, nearly resonant with the $S_{1/2}-P_{1/2}$ transition, in combination with another beam at 866~nm which is used to return population from the $D_{3/2}$ level into the cooling cycle. Pulses from a beam at 729~nm, resonant with the $S_{1/2}-D_{5/2}$ transition, coherently populate the $D_{5/2}$ level and are employed for sideband cooling and motional-state characterization. Another beam at 854~nm is responsible for deshelving population from the $D_{5/2}$ to the $S_{1/2}$ level via the intermediate $P_{3/2}$ level (see Fig.\ref{fig:diagram}a) \cite{roos_quantum_1999}. We distinguish between an ion in the $S_{1/2}$ level (bright) and one in the $D_{5/2}$ level (dark) via observation of state-dependent 397 nm fluorescence as recorded on a photomultiplier tube. After many experimental repetitions, the bright and dark state populations are estimated via maximum likelihood (see Supplemental Material of Ref. \cite{clark_2021}).

To characterize the ion's position throughout the full arc of its roughly linear trajectory, we implement a position-sensitive optical deshelving technique~\footnote{The use of a single optical deshelving laser beam does not allow us to directly distinguish between different trajectories perpendicular to the laser beam propagation direction. Additional orthogonal beam orientations would permit a full characterization of the three-dimensional ion trajectory}. For this, we prepare the ion in the $D_{5/2}$ state before transport and then apply a pulse (200~ns, shorter than the transport duration) of 854~nm light at a later time, thereby deshelving the $D_{5/2}$ level with a probability dependent on the laser beam's local intensity. By choosing the Gaussian waist $w_0$ of the 854~nm beam such that its intensity varies significantly and monotonically along the ion's trajectory ($w_0\sim100~\mu$m), we achieve a position-dependent deshelving probability ($P_d$). We can therefore invert spatially monotonic measurements of this probability, acquired at various times after the start of the transport, to determine the ion's position at these instants. The beam is directed perpendicular to the linear transport axis to remove velocity dependence (first-order Doppler shifts, e.g.) from the deshelving probability.

The above technique requires calibration of the deshelving probability for a given ion location using a procedure diagrammed in Fig.~\ref{fig:diagram}b. For this, we Doppler and sideband cool the ion nearly to its axial motional ground state $(\bar{n}<1)$ and prepare it in $S_{1/2}$. We then shelve it into the $D_{5/2}$ level with a resonant 729~nm $\pi$-pulse and subsequently move it adiabatically to a given position along the trap axis (as determined from an electrostatic model of the trapping potential). We illuminate the ion with a 200~ns, 854 nm deshelving pulse. After adiabatically returning the ion to its initial location, we determine its state by collecting fluorescence. Repeating this experiment 400 times each at 2~$\mu$m intervals yields a measurement of deshelving probability $P_d(z)$ for each location as shown in Fig.~\ref{fig:diagram}c. We fit $P_d(z)$ with a polynomial (red curve of Fig.~\ref{fig:diagram}c) rather than with a Gaussian to allow for possible distortion of the 854~nm laser beam mode shape, and we ensure that $P_d(z)$ is single-valued by having previously displaced the beam such that its intensity varies monotonically through the calibration region. 

It is important that any movement of ions from one place to another in the trap be accomplished with only a small degree of motional excitation. For the experiments outlined below, we design the waveform as a simple linear time interpolation of the potential minimum through a 120~$\mu$m displacement. We choose a transport duration (1.2~$\mu$s) that is an integer multiple of the ion's harmonic motional period (0.5~$\mu$s, nominally held constant during the transport). In the absence of filter distortion, such a waveform is expected to coherently excite ion motion at the beginning of transport but subsequently to suppress this excitation at the end \cite{Bowler2012, todaro_scalable}. Our filter bandwidths lie below the ion motional frequencies by design. At slow enough speeds, the action of the filters is to smooth out the beginning and ending accelerations, so that the ion's motion is not excited as much during transport as would otherwise be the case. At faster speeds such as are used here, filter distortion is great enough that simple scaling of the waveform confinement strength can no longer fully suppress the final excitation \cite{todaro_scalable}. However, provided that the waveform is performed identically for each experimental repetition and that the confinement remains harmonic, the final excitation corresponds to a coherent displacement. We remove it by superimposing on the transport waveform a sinusoidal rf pulse applied to four of the trap electrodes, and we optimize the phase, frequency, and amplitude of this pulse to achieve minimum mode occupation. 

\begin{figure}
    \centering
	\includegraphics[width=0.47\textwidth]{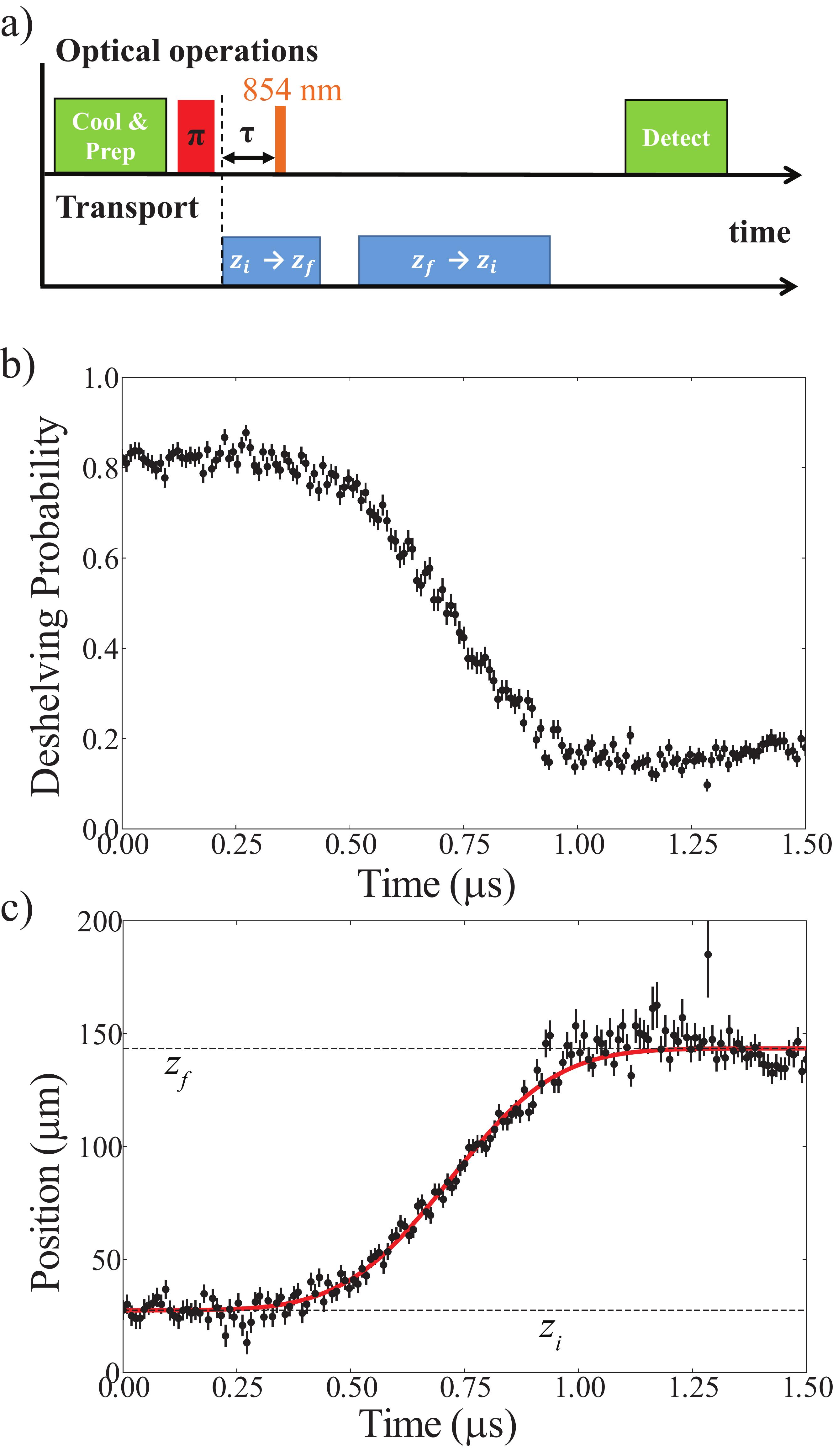}%
	\caption{\label{fig:velocity210} (color online)  (a) Ion trajectory measurement sequence. An ion in a stationary potential at $z_i$ is cooled and prepared in the $S_{1/2}$ level and then shelved into the $D_{5/2}$ level with a resonant 729 nm $\pi$-pulse. The transport waveform to $z_f$ begins to play, and following a delay $\tau$ the 854 nm beam is pulsed for 200 ns, thereby deshelving a portion of the $D_{5/2}$ population. The ion is shuttled adiabatically back to its initial location $z_i$ for final detection. (b) Experimental time-dependence of deshelving probability sampled at 10~ns intervals after the start of the fast transport waveform; error bars represent the 68\% confidence interval in state populations assuming binomial statistics. (c) Time-dependence of ion position, with experimental data (black points) and empirical fit (red line, see main text). Here, the experimental points are obtained by inverting the polynomial curve in Fig.~\ref{fig:diagram}c with the measured points in (b). The fit yields a maximum speed of \speed{}, while the waveform was designed with 30 samples at 40 ns intervals to shuttle the ion across a 120~$\mu$m displacement in 1.2~$\mu$s (\designSpeed{}).}
\end{figure}

With the $P_d(z)$ calibration complete, we then transport the ion through the full trajectory of 120~$\mu$m. Fig.~\ref{fig:velocity210}a gives a diagram of the sequence. We pulse the 854~nm deshelving laser for 200~ns while the ion is in motion, with a configurable delay between the start of the transport waveform and the start of the pulse. The ion is then returned adiabatically to its initial location for state detection~\footnote{Final state detection is independent of the return speed after deshelving provided that the ion is not sufficiently excited during the return to modify the its fluorescence.}. With 100 repetitions of this experiment at each delay
, we obtain a map $P_d(t)$ of deshelving probability in time following the start of the waveform (Fig.~\ref{fig:velocity210}b).
To obtain the corresponding position map $z(t)$, we invert the polynomial fit of Fig.~\ref{fig:diagram}c to obtain $z(P_d)$ and then compute $z(P_d(t))$, producing the points in Fig.~\ref{fig:velocity210}c. Finally, we fit the $z(t)$ data of Fig.~\ref{fig:velocity210}c with the following phenomenological expression to extract the mean and maximum linear velocities:

\begin{equation} \label{eq:position_erf}
    z(t) = z_i + \frac{\sqrt{\pi}}{2} v_\mathrm{max} t_{\sigma} \left[ \mathrm{erf}\left(\frac{t-t_c}{t_{\sigma}}\right) - \mathrm{erf}\left(\frac{t_0-t_c}{t_{\sigma}}\right) \right].
\end{equation}

Equation~\ref{eq:position_erf} is derived assuming that the speed follows a Gaussian profile in time, an empirical assumption justified by its agreement with the $z(t)$ data in Fig.~\ref{fig:velocity210}c. Here, $z_i$ represents the initial ion position, $t_0$ is the initial time (time when $z(t)=z_i$), $t_{\sigma}$ is the $1/e$ temporal half-width of the ion speed, and $t_c$ is the time of maximum speed. We use the following standard definition for the error function:

\begin{equation}
    \mathrm{erf}(x) = \frac{2}{\sqrt{\pi}} \int_0^x e^{-y^2}dy.
\end{equation}

We note that a naive estimate of the mean speed, obtained from the 120~$\mu$m and 1.2~$\mu$s waveform displacement and duration, would be \designSpeed{}. In contrast, the fit of Fig.~\ref{fig:velocity210}c (red curve) yields a much higher maximum slope of \speed{}. To determine an effective mean speed, we estimate the beginning and ending times of the transport by determining when the ion is within a given distance of its asymptotic positions. Such a choice must always be made if we are to take into account the influence of the electrode filters on the results, just as similar cutoffs must be chosen when studying the time response of such analog filters more generally. In particular, here we choose a distance from the fitted asymptotes that equals the ground-state extent of the 2.2~MHz trap potential, approximately 8~nm. This rather arbitrary decision, as well as our selection of Eq.~\ref{eq:position_erf} as a model for ion position, both play an outsized role in our determination of the mean ion speed and highlight the need to define these terms with sufficient detail in studies of ion transport. With these choices we determine a mean speed of \averageSpeed{}, slightly lower than the naive estimate.

To avoid additional gate errors within a quantum algorithm, fast transport must not excite excessive ion motion. For trapped ions in thermal states of motion probed in the Lamb-Dicke regime, one can measure the ratio of the first red and first blue sideband excitations to determine the mean thermal mode occupation ($\nth$)~\cite{turchette_2000}. However, characterization of non-thermal (e.g. coherent) distributions is more complicated since the ratio of first sidebands can vary with the probe duration. Given that fast transport can leave the ion with a large coherent excitation ($\ncoh$)~\cite{Bowler2012, walther_controlling_2012}, we directly fit the time-dependent excitation of the first blue sideband assuming a convolution of coherent and thermal distributions~\cite{walther_controlling_2012}. 

To measure the ion's axial motional excitation after the full transport (diagram in Fig. \ref{fig:diagram}b), we apply the same waveform as before but do not shelve the ion with an initial $\pi$-pulse. Instead, following the ion's adiabatic return to its initial position, we drive the blue axial motional sideband of the $S_{1/2}-D_{5/2}$ transition and we analyze the dependence of $S_{1/2}$ state populations $P_S$ on pulse duration (sideband flopping curves). This probability is sensitive to the motional state and yields information about both the average mode occupation and its statistical distribution~\cite{meekhof_generation_1996}. Neglecting the radial modes, it is given by

\begin{equation}\label{eq:blueflop}
P_{S}(t) = \frac{1}{2} \left(1+e^{-\gamma t}\sum_{n=0}^{\infty} p_n\cos(2\Omega_{n,n+1}t)\right)
\end{equation}
where $p_n$ is the mode population fraction in the Fock state $\ket{n}$, $\gamma$ is a phenomenological decoherence rate, and $\Omega_{n,n+1}$ is the Rabi frequency for the first blue sideband transition for an ion starting in $\ket{n}$. We use the full expression for the first blue sideband Rabi frequency \cite{wineland_experimental_1998},
\begin{equation}
    \Omega_{n,n+1} = \eta \Omega_0 e^{-\eta^2 / 2} \sqrt{\frac{1}{n+1}} L_n^1(\eta^2),
\end{equation}
where $\Omega_0$ is the optical carrier Rabi frequency, $\eta$ is the Lamb-Dicke parameter, and $L_n^1$ is the $n^{\text{th}}$ associated Laguerre polynomial of order 1. As a simple model, we assume that any excitation can be represented as a convolution of thermal and coherent contributions \cite{meekhof_generation_1996,walther_controlling_2012}, and we fit the measured probabilities to Eq.~\ref{eq:blueflop} under this assumption.

\begin{figure}
    \centering
	\includegraphics[width=0.48\textwidth]{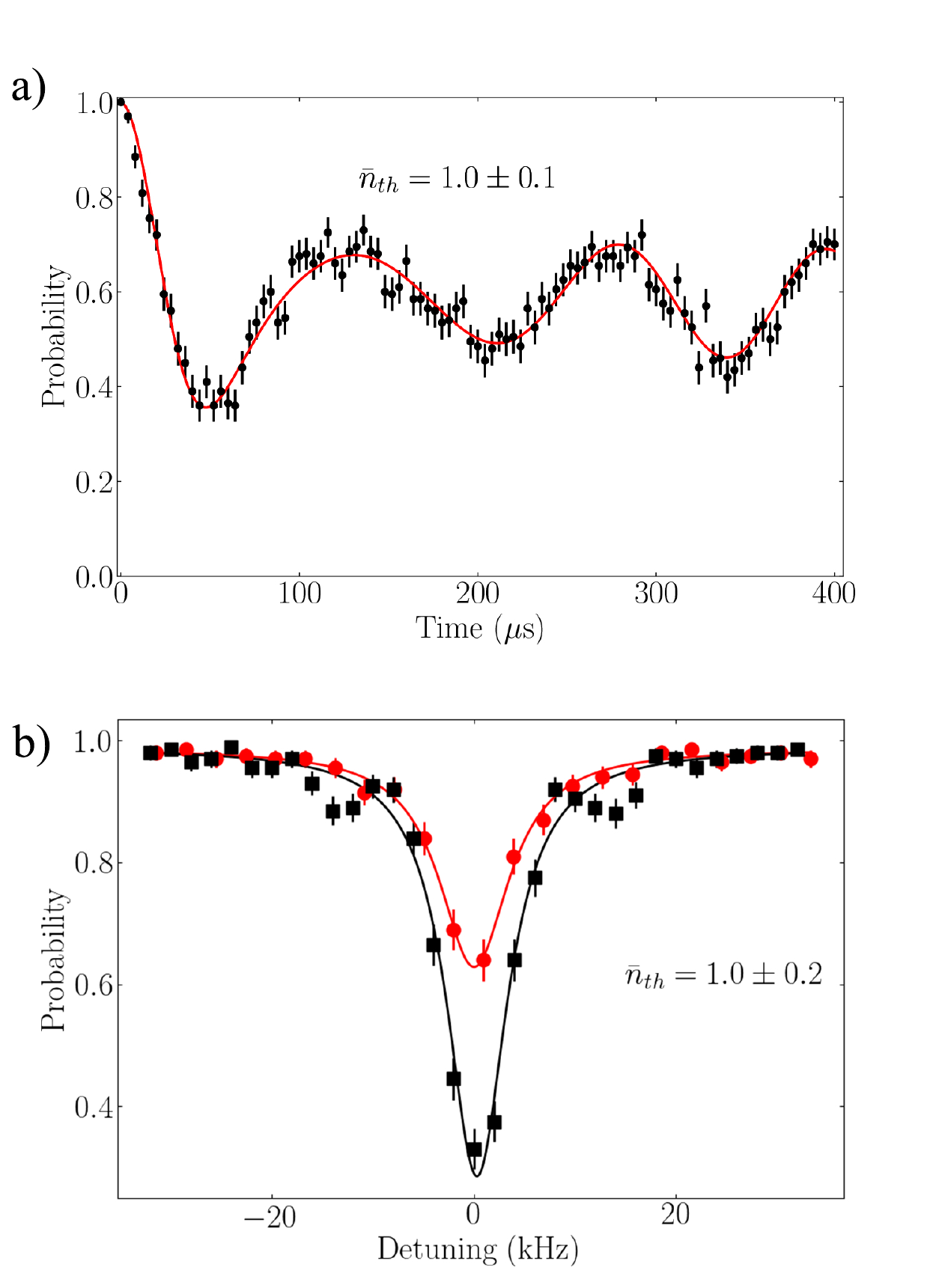}
	\caption{\label{fig:fastTransportdata} (color online) Characterization of ion temperature after fast transport. (a) We perform a pulse on the axial blue motional sideband for a variable duration after the fast transport operation described in Fig.~\ref{fig:velocity210}. The red trace represents a fit of the data (black points) to Eq.~\ref{eq:blueflop}, which yields a purely thermal axial mode excitation of 1.0(1) quanta. Here we have added to the transport waveform a sinusoidal oscillation near the ion axial frequency with appropriate amplitude and phase to remove transport-induced coherent excitation. (b) Red (red circles) and blue (black squares) sideband lineshapes, also measured after optimized fast transport, confirm that the ion is nearly in the ground state. Fits (solid curves) to the data (individual points) confirm the low ion temperature: the ratio of sideband amplitudes corresponds to 1.0(2) quanta. Error bars represent the 68\% confidence interval in state populations assuming binomial statistics}
\end{figure}

Figure~\ref{fig:fastTransportdata}a shows a fit to the time-dependent blue sideband excitation, revealing a purely thermal ($\ncoh = 0$) excitation of $\nth=1.0(1)$. Having determined that the excitation is thermal, we verify the mode temperature through a comparison of red- and blue-sideband transition amplitudes \cite{turchette_2000} (Fig.~\ref{fig:fastTransportdata}b). This yields a post-transport temperature of $\nth=1.0(2)$ compared to $\nth=0.3(1)$ measured before transport, and we conclude that the optimized transport induces an additional 0.7(2) quanta of motional excitation. We note that, without the resonant de-excitation of motion during the transport operation, we measure an additional \textit{coherent} excitation of $\ncoh = 61.7(6)$.
Resonant de-excitation can be extended to the collective modes of a multi-ion crystal provided that electric fields with sufficiently high spatial frequency can be generated.

In conclusion, we have developed a general method for experimentally characterizing ion transport trajectories using position-dependent optical deshelving, and we verified the technique in a surface-electrode ion trap by shuttling an ion along a linear trajectory of 120~$\mu$m (two electrode widths) with a 1.2~$\mu$s waveform. Owing to the impact of filter distortion on the transport potentials, the ion reaches instantaneous speeds significantly higher than might be naively assumed from the waveform design. We characterized the final motional state using two complementary methods to fit blue sideband flop curves as well as red and blue sideband lineshapes. Even at this high speed the transport incurs only 0.7(2) quanta of axial excitation, small enough to have minimal impact within a quantum algorithm.  


Beyond single-ion transport through linear sections, this technique could also be applied to optimize fast merging and separation of ions into chains \cite{Bowler2012, ruster_2014}. With the incorporation of multiple deshelving wavelengths, the positions of disparate ion species could be tracked simultaneously \cite{pino_2021}. The method might prove particularly useful when optimizing the paths of ions through junctions of linear sections, where trajectories deviate significantly from straight lines both horizontally and vertically. With multiple deshelving beams at complementary angles one could isolate an ion's position in all three dimensions. 

This work was done in collaboration with Los Alamos National Laboratory.

\bibliography{transport}

\end{document}